# Dynamical Anisotropic Response of Black Phosphorus under Magnetic Field

Xuefeng Liu[1,2], Wei Lu[1,2], Xiaoying Zhou[3,4], Yang Zhou[3], Chenglong Zhang[1,2], Jiawei Lai[1,2], Shaofeng Ge[1,2], Chandra Sekhar Mutyala[1,2], Shuang Jia[1,2], Kai Chang[3,*] and Dong Sun[1,2,*]

[1]International Center for Quantum Materials, School of Physics, Peking University, Beijing 100871, P. R. China

[2]Collaborative Innovation Center of Quantum Matter, Beijing 100871, P. R. China

[3]SKLSM, Institute of Semiconductors, Chinese Academy of Sciences, P.O. Box 912, Beijing, 100083, China

[4]Department of Physics and Key Laboratory for Low-Dimensional Quantum Structures and Manipulation (Ministry of Education), and Synergetic Innovation Center for Quantum Effects and Applications of Hunan, Hunan Normal University, Changsha 410081, China

[*]Email: sundong@pku.edu.cn (D.S.); kchang@semi.ac.cn (K.C.)

**Abstract:**

Black phosphorus (BP) has emerged as a promising material candidate for next generation electronic and optoelectronic devices due to its high mobility, tunable band gap and highly anisotropic properties. In this work, polarization resolved ultrafast mid-infrared transient reflection spectroscopy measurements are performed to study the dynamical anisotropic optical properties of BP under magnetic fields up to 9 T. The relaxation dynamics of photoexcited carrier is found to be insensitive to the applied magnetic field due to the broadening of the Landau levels and large effective mass of carriers. While the anisotropic optical response of BP decreases with increasing magnetic field, its enhancement due to the excitation of hot carriers is similar to that without magnetic field. These experimental results can be well interpreted by the magneto-optical conductivity of the Landau levels of BP thin film, based on an effective $k \cdot p$ Hamiltonian and linear response theory. These findings suggest attractive possibilities of multi-dimensional controls of anisotropic response (AR) of BP with light, electric and magnetic field, which further introduces BP to the fantastic magnetic field sensitive applications.

As a two-dimensional (2D) tunable-gap semiconductor with high mobility[1-7], black phosphorus (BP) promises enormous applications in high speed transistors with large on-off ratio[7-13] and mid-IR optoelectronic devices[14-19] surpassing graphene and many other two-dimensional materials[20-23]. Its highly anisotropic response (AR) would lead to interesting transport and optical properties, such as anomalous magneto-optical response[24], highly anisotropic wave-functions of the Landau levels (LLs)[25], and quasi-one dimensional excitonic state with larger binding energy[26, 27], which enables an additional degree of freedom for high performance functional device applications, especially in the applications where angle sensitive operations are desired[3, 5, 26, 28-31]. Investigating the AR of BP under various circumstances such as high field/speed limit or magnetic field are crucial for BP based angle sensitive devices. The tuning with magnetic field is greatly facilitated by recent introduction of 2D Ferromagnetic materials[32-34] and convenient integration of van der Waals heterostructures[35]. Under these prospectives, steady state and magneto transport, optical characterization[26, 28, 31] and even high speed transport properties[36] have been applied on BP to characterize its fundamental anisotropy related physical responses. However, despite its importance in fundamental physics and device applications, the dynamical anisotropic optical response of photoexcited carriers under magnetic field, which is vital for any magnetic field related applications, has not been investigated experimentally so far.

In this work, polarization resolved ultrafast mid-infrared transient reflection spectroscopy measurements are performed under magnetic fields up to 9 T to study the dynamical evolution of anisotropic optical properties of BP. The relaxation dynamics of photoexcited carrier, which can be directly correlated to the transient reflection measurements, are found to be insensitive to the applied magnetic field up to 9 T due to the broadening of the LLs and large effective mass of carriers. Through pump polarization dependent magneto-transient reflection measurements, the AR of BP at pump photon transition can be deducted, and it is revealed from these measurements that the anisotropic optical response of BP decreases with increasing magnetic field. By performing the calculation of the magneto-optical conductivity of the LLs of black phosphorus thin film, based on an effective $k \cdot p$ Hamiltonian and linear response theory, the results indicate the degradation of the AR is an effect of the compression of electron wave-function by magnetic fields. On the other hand, the dynamical evolution of optical conductivity at probe photon transition due to the excitation of hot carriers can be resolved in probe polarization dependent magneto-transient reflection measurements. The results indicate the enhancement of the AR of BP due to the excitation of hot carriers persists under magnetic fields, which is similar to our previous work on the AR of BP under high field/speed operation limit but without magnetic field.[36]. Further analysis indicates the relative enhancement of the AR due to hot carrier excitations does not decrease with increasing magnetic field, thus the hot-carrier acceleration either with photoexcitation or high electric and applying a magnetic field can provide multiple dimensional control of BP's anisotropic degree of freedom.

The schematic diagram of transient magneto-optical measurements is shown in Figure 1a and more experimental details and basic BP sample characterizations are described in the methods section. The effect of magnetic field on the AR of BP is determined by transient measurements with two different configurations of pump/probe wavelengths: 800 nm/1940 nm and 1940 nm/4000 nm with the magnetic field applied perpendicular to the BP xy-plane. Figure 1b and 1c show the optical and atomic force microscopy images of a typical BP sample that is mechanically exfoliated from synthesized bulk BP crystal. The x- and y-axes are along the armchair and zigzag direction, respectively, which is identified by the polarization-resolved reflection spectroscopy described in reference[36]. Thickness of this BP sample measured by atomic force scanning along the white dotted line shown in Figure 1d is about 101 nm, from which we can deduce that the sample contains around 190 layers, using a layer spacing of 0.53 nm. The synthesized BP crystal is found to be p-doped with bulk doping density of $4.2\times10^{18}$ cm$^{-3}$ by Hall measurement, which converts to $4.24\times10^{13}$ cm$^{-2}$ for 101-nm thick BP flake. The typical pump fluence for 800-nm and 1940-nm excitation is $8.1\times10^{14}$ and $5.4\times10^{15}$ photons/cm$^2$ respectively, which converts to $3.27\times10^{14}$ and $2.18\times10^{15}$ cm$^{-2}$ electron-hole pairs after the excitation assuming a 0.4%/nm linear absorption rate at both wavelengths[2]. Figure 1e shows the band structure of bulk BP and the difference of dispersion relation along Γ-X and Γ-Y direction indicating that the effective masses are also anisotropic. The probe photon energy, either 1940 nm or 4000 nm, involves the transition from the highest valence band and lowest conduction band only. To measure the dynamic evolution of the AR, polarization resolved 1.55 eV (0.8-μm) /0.64 eV (1.94-μm) and 0.64 eV (1.94-μm)/0.31 eV (4-μm) probe photon arriving at various delay time (t) with respect to pump pulse is used and the pump induced probe reflection change is recorded.

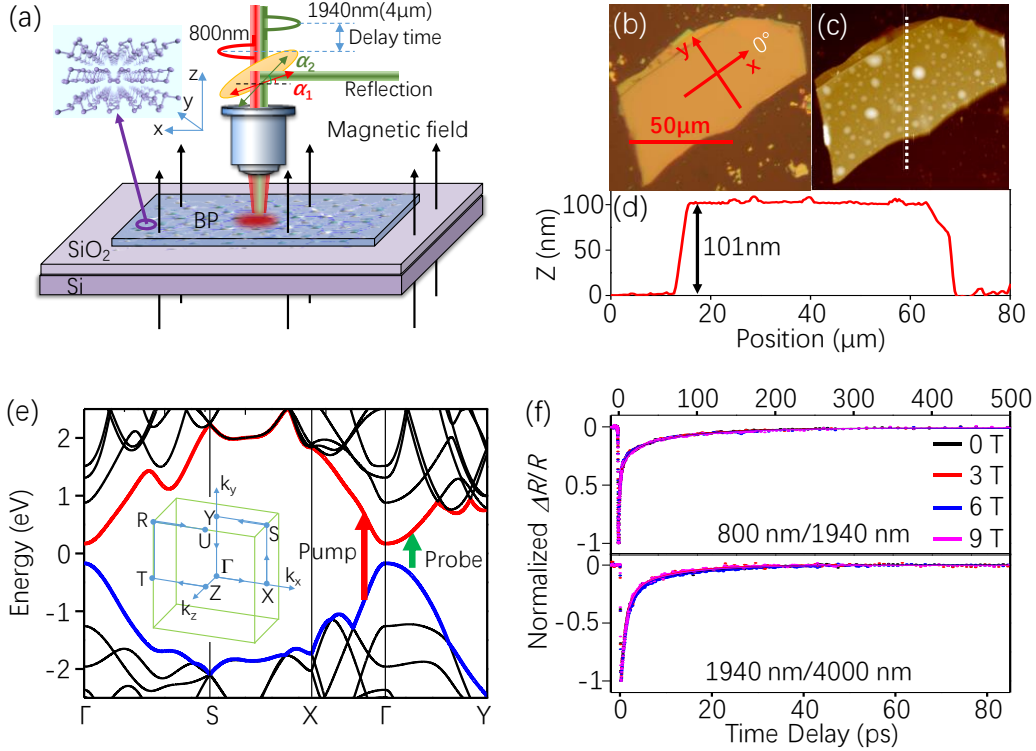

**Figure 1. Experimental scheme, sample characterization and representative transient reflection spectra. a,** Schematic diagram of polarization-resolved transient reflection experimental setup under magnetic field. **b,** Optical image of a BP sample in which the crystal x, y axes are marked by red arrows, respectively, and **c.** atomic force micrograph of the BP sample, **d.** Atomic force scanning of sample thickness along the white dotted line in **c**. **e.** Band structure of bulk BP and inset shows the Brillouin zone. **f.** Transient reflection kinetics under 0, 3, 6, and 9 T magnetic field at 78 K with pump/probe wavelength of 800 nm/1940 nm (upper panel) and 1940 nm/4000 nm (lower panel) respectively. In both cases, the pump polarization is along x-direction. The solid lines are fittings by biexponential function.

## Results

**Photoexcited carrier dynamics under magnetic field.**

Figure 1f shows typical magneto-transient reflection kinetics with pump polarization set along crystal x-axis and probe polarization either parallel or perpendicular to that of the pump. In either cases, the transient reflection signals $\Delta R/R$ are negative at timezero and then the signal evolves differently with different probe polarizations, which has been discussed extensively in our previous work[36]. Interestingly, in the present study, it is observed that the perpendicular magnetic field does not have any influence on the dynamical evolution of $\Delta R/R$ of BP up to 9 T with either 800-nm/1940-nm or 1940-nm/4000-nm pump probe configurations. Usually, the discrete LLs that are induced in high magnetic fields will increase the photoexcited carrier relaxation time compared to continuous band, as the relaxation between discrete energy level has to fulfill more rigorous energy and momentum scattering process[37]. However, the observed dependence of the dynamic evolution of $\Delta R/R$ on magnetic field (up to 9 T) indicates that the photoexcited carrier relaxation time of BP is not influenced by the magnetic field.

**Degradation of anisotropic response under magnetic field.**

Though the magnetic field does not influence the photoexcited relaxation time of BP, it has a profound impact on the wave-function of BP and thus affects its AR. Effect of magnetic field on the AR of BP can be determined by measuring the transient reflection amplitude at different pump polarization as shown in Figure 2. It is clearly evident from Figure 2 that the amplitude of transient reflection $\Delta R/R$ varies with pump polarization, while the shape of their kinetics remains the same (Figure 2a). In our previous work, it has been shown that the amplitude of transient reflection at timezero ($\Delta R/R|_{t=0}$) has linear dependence on photoexcited carrier density based on the pump power dependent measurement[36] and the $\Delta R/R|_{t=0}(\alpha_1)$ is proportional to pump polarization angle ($\alpha_1$, respective to crystal x-axis as marked in Figure 1a) dependent absorption coefficient $A(\alpha_1)$. Based on this, we can simply fit $\Delta R/R|_{t=0}(\alpha_1)$ as function of $\alpha_1$ (see Supplementary Note 1 for derivation):

$$\Delta R/R|_{t=0}(\alpha_1) \sim A(\alpha_1) \sim \frac{4\sqrt{\varepsilon_1}(Re(\sigma_{xx})cos^2\alpha_1 + Re(\sigma_{yy})sin^2\alpha_1)}{\varepsilon_0 c(\sqrt{\varepsilon_2}+\sqrt{\varepsilon_1})^2} \quad (1)$$

where, $\varepsilon_1$ and $\varepsilon_2$ are dielectric constants of substrate and air, respectively, $\varepsilon_0$ is the free-space permittivity, $c$ is the speed of light, $\sigma_{xx}$ and $\sigma_{yy}$ are non-zero diagonal component of optical conductivity tensor. According to equation (1), $\Delta R/R|_{t=0}(\alpha_1=0°) \sim Re(\sigma_{xx})$, $\Delta R/R|_{t=0}(\alpha_1=90°) \sim Re(\sigma_{yy})$, so a measurement of $\Delta R/R|_{t=0}(\alpha_1)$ as function of $\alpha_1$ directly provides the conductivity ellipse $\sigma(\alpha_1)$ of BP at optical transitions corresponding to pump photon energy. In order to identify the influence of magnetic field on $Re(\sigma_{xx})$ and $Re(\sigma_{yy})$, pump polarization dependences under different magnetic fields are investigated experimentally. Figure 2b shows the evolution of experimentally measured conductivity ellipses as function of applied magnetic fields. We note although both $Re(\sigma_{xx})$ and $Re(\sigma_{yy})$ increases with applied magnetic field, the relative increment $Re(\sigma(B))/Re(\sigma(0))$ is much larger in y-direction compared to that in x-direction (Figure 2c). The ratio $\frac{Re(\sigma_{xx})}{Re(\sigma_{yy})}|_{1940\ nm}$, which describes the AR of BP at 1940 nm transition, decreases from 4.1 at 0 T to 2.7 at 9 T (Figure 2d). Similar trends are also observed for the AR of BP at 800-nm pump; $\frac{Re(\sigma_{xx})}{Re(\sigma_{yy})}|_{800\ nm}$ decreases from 6.8 at 0 T to 2.7 at 9 T (see, Supplementary Fig. 3). These results clearly indicate that the AR of BP degrades significantly with increasing magnetic fields. This phenomenon can be understood from a semi-classical point of view. Under a magnetic field, charged carriers will do cyclotron motions. During this process, the wave-functions with momentums along different directions will be mixed, and hence the optical AR of BP tends to become less anisotropic. More information can be found in the discussion part and the theoretic details are presented in Supplementary Note 8.

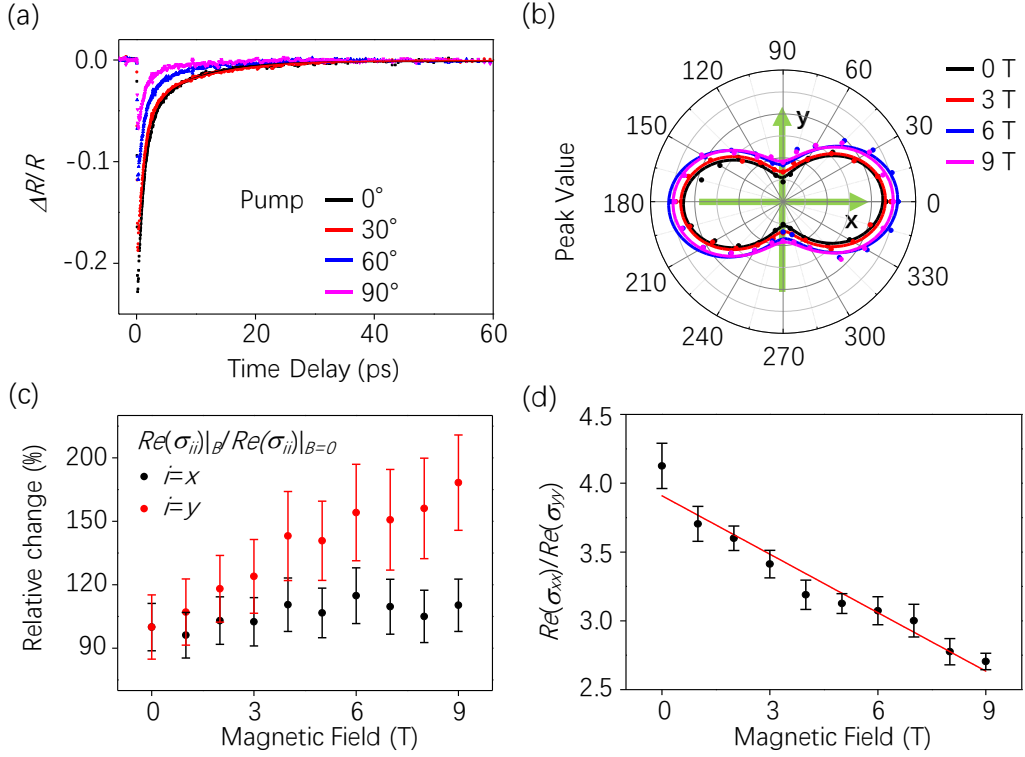

**Figure 2. Evolution of AR under magnetic field. a,** Transient reflection kinetics with 1940-nm pump polarization fixed at 0°, 30°, 60° and 90° with respect to x-axis and 4000-nm probe polarization fixed along x-axis. The solid lines are biexponential fittings. **b,** $\Delta R/R|_{t=0}$ as a function of pump polarization with magnetic field of 0, 3, 6 and 9T respectively. The polarization dependence of $\Delta R/R|_{t=0}$ is fitted by the function $A_1 cos^2(\alpha_1+\varphi_B)+A_2 sin^2(\alpha_1+\varphi_B)$. **c,** The relative change of real part of conductivity $Re(\sigma_{ii})|_B/Re(\sigma_{ii})|_{B=0}$, $i=x$ or $y$. The black and red circles represent conductivity change in x- and y-axis, respectively. **d,** The anisotropy of conductivity ellipse: $Re(\sigma_{xx})/Re(\sigma_{yy})$ as a function of magnetic field. The red solid line is a linear fitting. All experiments are taken at 78 K.

**Magnetic field enhanced relative anisotropy increment with hot carriers.**

Furthermore, the hot carrier induced anisotropy change under magnetic field is also investigated from probe polarization dependent transient reflection measurements. In the absence of magnetic field, the photoexcited hot carriers can enhance the AR of BP, which has been systematically studied in our previous work[36]. Figure 3a shows the probe polarization dependence of transient reflection kinetics under 9 T magnetic field. The dynamical evolution of the transient reflection signal with different probe polarization angle is similar to the one in the absence of magnetic field. At fixed pump-probe delay, $\Delta R/R$ demonstrates sinusoidal oscillation as a function of probe polarization (Figure 3b). Except for the initial non-equilibrium state, this periodic oscillation is robust and persists during the whole decay process (Figure 3b). After the initial relaxation of highly excited states, the measured $\Delta R/R$ can be converted to optical conductivity change $\Delta Re(\sigma)$ through the following relation (see Supplementary Note 4 for derivation):

$$\frac{\Delta R}{R}\big|_t(\alpha_2) \approx \frac{0.03}{\sigma_0}(\Delta Re(\sigma_{xx})cos^2\alpha_2 + \Delta Re(\sigma_{yy})sin^2\alpha_2) \qquad (2)$$

where $\alpha_1$ is probe polarization angle respective to crystal x-axis as marked in Figure 1a. As can be seen in Figure 3d, under high magnetic field at 9 T, the conductivity ellipse is stretched at pump-probe delay around 5 ps and later on gradually recovers as the delay time between pump and probe increases. The stretch of conductivity ellipse is similar to that in the absence of magnetic field, which is a result of photoexcitation of hot carriers. This behavior persists over the subsequent relaxation process and completely recovers after the relaxation and recombination of the photoexcited hot carriers. We can conclude from these results that the enhancement of ARin presence of hot carrier is well preserved under high magnetic field.

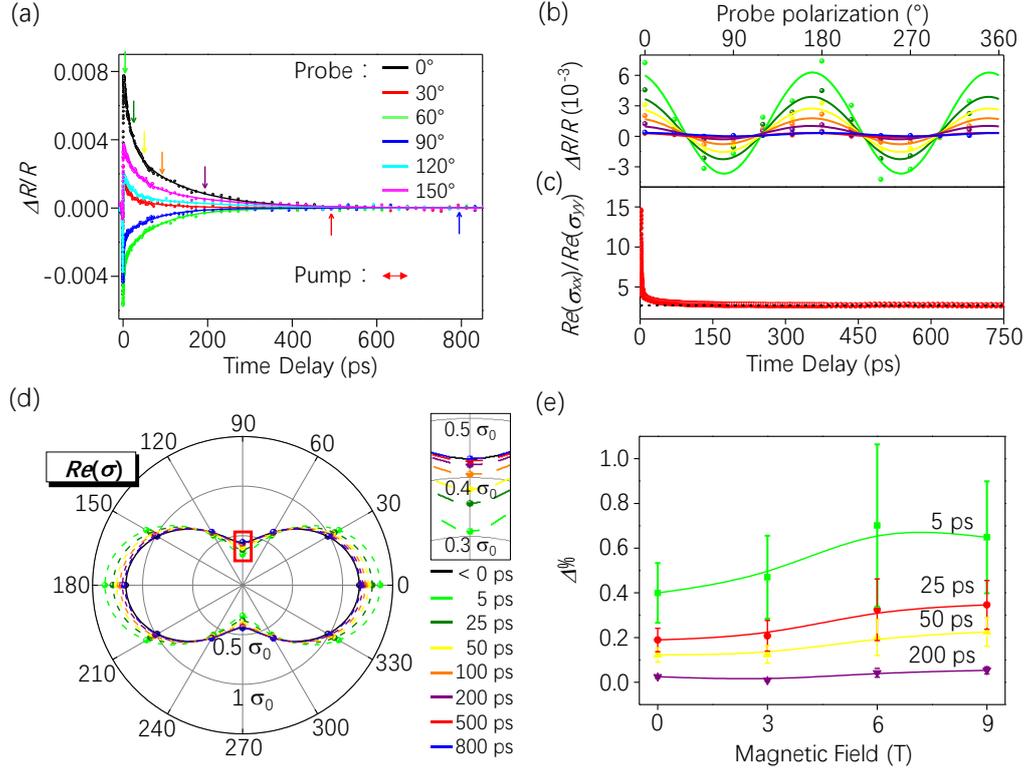

**Figure 3. Hot carrier induces anisotropy change under magnetic field. a,** 1940-nm probe polarization dependence of transient reflection kinetics with 800-nm pump polarization fixed along x-axis under 9 T magnetic field. **b,** Transient reflection spectra with different probe polarization at fixed delays marked in Figure **a**. The angle dependence is fit by function $\Delta R/R = acos^2\alpha_2 + bsin^2\alpha_2$. **c,** Dynamic evolution of $Re(\sigma_{xx})/Re(\sigma_{yy})$ at different delays at 9 T. The black dash line marks value of $Re(\sigma_{xx})/Re(\sigma_{yy})$ prior to pumping at 9 T. **d,** Dynamical evolution of $Re(\sigma)$ ellipse at different delays under 9 T magnetic field. $\Delta Re(\sigma)$ is doubled for clarity. Inset: Enlarged plot of area marked by red rectangle. **e,** The relative change $(\Delta\%)$ of anisotropy induced by photoexcited hot carriers at different magnetic fields with fixed pump-probe delay times. All experiments are performed at 78 K.

To analyze the magnetic field dependence of hot carriers enhancement on AR, we further analyze the enhancement of AR ($\Delta(B) = \frac{Re(\sigma_{xx})+\Delta Re(\sigma_{xx})}{Re(\sigma_{yy})+\Delta Re(\sigma_{yy})}\big|_B - \frac{Re(\sigma_{xx})}{Re(\sigma_{yy})}\big|_B$) due to the presence of hot carriers

at fixed delay as a function of magnetic field. To account for the degradation of anisotropy under magnetic field, the relative change $\Delta\%(B) = \Delta(B)/(\frac{Re(\sigma_{xx})}{Re(\sigma_{yy})}|_B)$ is plotted in Figure 3e. We note that $\Delta\%$ increases as magnetic field increases at all fixed pump-probe delays. Although the effect is not unambiguous considering the larger synthetic error bar of $\Delta\%$, however, it is safe to conclude that the magnetic field effect obviously does not degrade the relative enhancement of AR due to the presence of hot carriers.

**Discussion**

The experimentally observed magnetic field effect on dynamic evolution of photoexcited carriers can be phenomenological described by calculating the LL splitting and density of states around the probe photon transitions. To account for the temperature and inhomogeneous broadening of LLs in the sample, an anisotropic two-band *k·p* Hamiltonian of BP coupled with a perpendicular magnetic field B and a Gaussian broadening Γ of the LLs is incorporated into the calculation. The details of this calculation are described in the Supplementary Note 8. The LLs without the broadening which get involved into the 1940-nm and 4000-nm probe transition with $k_z=0$ are shown in Figure 4(a). Due to the large effective mass, the LLs are quite dense, and it's splitting at probe transition of 0.64 eV (1940 nm) and 0.31 eV (4000 nm) are only 4.3 (3.4) meV and 3.8 (2.8) meV, respectively, for valence (conduction) band under 9 T magnetic field. These tiny splittings can be smeared out by temperature (T=78 K, kT~6.72 meV) and inhomogeneous broadening due to disorder effect caused by impurities and defects. Figure 4(c) shows the density of states taking different LL broadening Γ. With Γ=1 meV the LL splittings are barely visible in 9 T magnetic field and the peaks of the density of states are completely quenched by LL broadening (Γ=4 meV, Figure 4d). Hence, the relaxation of photoexcited carriers from highly excited states to probe photon transition experiences continuous state instead of discrete Landau levels even under 9 T magnetic field, because of which the dynamics of photoexcited carrier relaxation (as observed in Figure 1f) is not influenced by the applied magnetic field.

The AR under magnetic field could be described approximately by magneto-optical conductivity (MOC) of BP. From the *k·p* Hamiltonian of BP coupled with a perpendicular magnetic field B and Kubo Formula[24], we can evaluate the longitudinal optical conductivities and plot the wave-function as functions of the photon energy and magnetic field (see, Supplementary Note 8 ). The spatial distribution of the wave-function for the LLs in the valence band at the excitation photon transition of 0.64 eV at $k_z=0$ are shown in Supplementary Fig. 7. Comparing to the wave-function at weaker magnetic field (B=6 T), the wave-function under 9 T magnetic field is less anisotropic. Quantitatively, if we plot the $Re(\sigma_{xx})/Re(\sigma_{yy})$ as a function of magnetic field which is shown in Figure 4b, $Re(\sigma_{xx})/Re(\sigma_{yy})$ gradually decreases with increasing magnetic field, which is consistent with experimental observation. However, there are two major discrepancies comparing with experimental results: First, $Re(\sigma_{xx})/Re(\sigma_{yy})$ oscillates with magnetic field in Figure 4b, which is not observed in experiment. Measurement of $Re(\sigma_{xx})/Re(\sigma_{yy})$ with fine tuning of magnetic field does not show clear oscillatory behavior (Supplementary Fig. 6), this is due to the quenching of the Landau level splittings at transitions corresponding to the pump photon energy as shown in Figure 4a. Second, the absolute magnitude of $Re(\sigma_{xx})/Re(\sigma_{yy})$ is far larger than those observed in the experiment. Theoretically, the $Re(\sigma_{xx})/Re(\sigma_{yy})$ highly depends on the temperature and doping intensity as shown in Supplementary Fig. 7. As the experiment is performed with femtosecond pulse, the absorption of falling edge of the pulse is strongly affected by the highly excited unequilibrium state of BP that are excited by the rising edge of the pulse, thus experimentally observed $Re(\sigma_{xx})/Re(\sigma_{yy})$ is strongly modified because the excited carriers and the magnitude is significantly different from our theory which is based on quasi-equilibrium state.

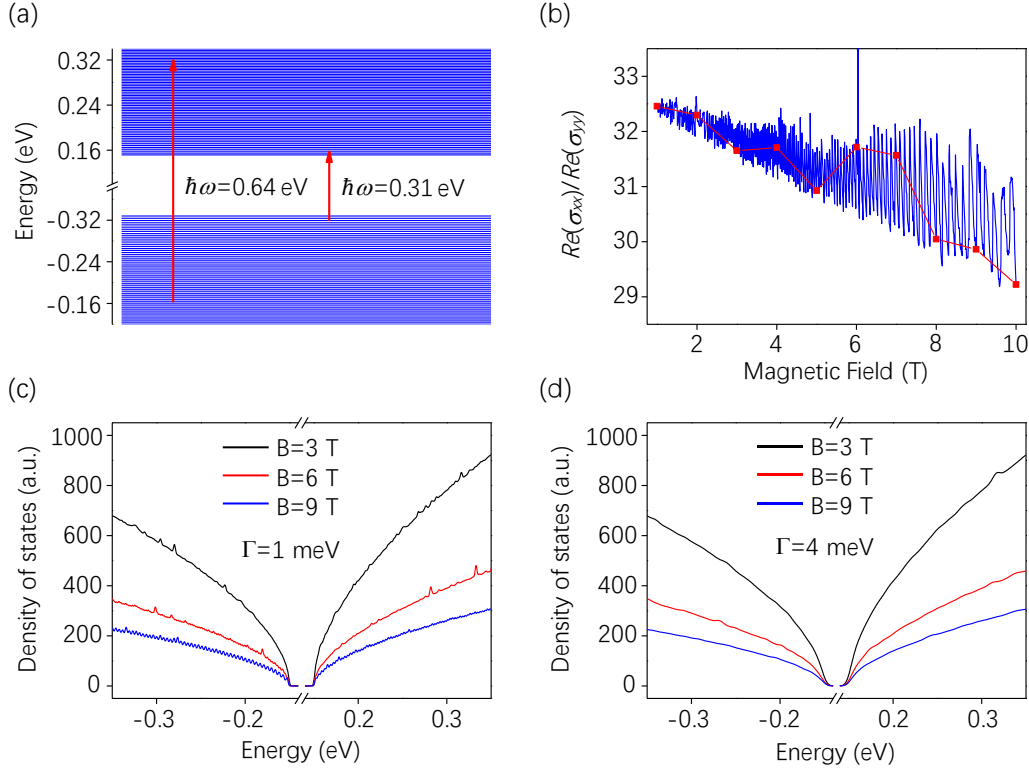

**Figure 4. Theoretical calculation of the AR under magnetic field. a,** Landau levels of black phosphorus under 9 T magnetic field with $k_z=0$. **b,** The ratio of the magneto-optical conductivity along the armchair (x) and zigzag (y) direction: $Re(\sigma_{xx})/Re(\sigma_{yy})$ as function of magnetic fields. The red dot marks the results at integer magnetic field. **c-d,** The density of states under different magnetic fields. The Landau level broadening Γ is adopted as 1 meV and 4 meV in **c** and **d**, respectively. The kinks in the density of states are arising from the crossing of the Landau levels from different sub-bands in z-direction.

Although the magnetic field significantly degrades the AR of BP, the hot carrier enhancement effect on AR is preserved under magnetic field as verified in probe polarization dependent measurement. These experimental results further extend the tuning capability on AR with hot carriers under magnetic field. Thus, in addition to magnetic field, excitation of hot carriers provides another tuning knob of the AR of BP which stays robust under high magnetic field. The hot carriers excitation with light or high electric field together with applying magnetic field provide multiple dimensional control of AR for high performance device with BP. Though the magnetic field control is usually considered not as convenient as electric field, recent introduction of two dimensional ferromagnetic semiconductor materials[32, 33] and the easy van der Waals integration with other layered 2D materials could greatly facilitate the magnetic field tuning of BP. Furthermore, large range bandgap tuning of few-layer BP by electrical gating has also been realized experimentally[38-40], this dynamic tuning of bandgap may further realize electrically tunable topological insulators and semimetals with BP[41-43]. Together with other tuning knobs, such as strain, pressure[44, 45], BP is becoming a very versatile materials platform that can be easily tuned between different condensed matter phases. Based on the above, the AR is adding another controllable degree of freedom to BP, which can be used for applications when angle sensitive application is desired. The experimental verification of the multiple dimensional control of this degree of freedom with high electric field acceleration of hot carriers and magnetic field will further pave the way for future multiple functional high performance devices based on BP.

**Methods:**

**Sample preparation.**
Bulk black phosphorus was grown by chemical vapor deposition as described before[46]. Red phosphorus (500 mg, 99.999+%), AuSn (364 mg) and $SnI_4$ were sealed in an evacuated quartz ampule of about 12-cm length and 1-cm diameter (pressures lower than $10^{-3}$ mbar). Then this ampule was placed horizontally at the center of a single zone tube furnace chamber. The ampule was slowly heated to 873 K in 10 hours and maintained at this temperature for 24 hours. In a next step the temperature was reduced to 773 K applying a cooling rate of 40 K/hour and maintained this temperature before switching off the oven. Using this method black phosphorus with crystal sizes larger than 1-cm can be obtained. The thin black phosphorus flake was mechanically exfoliated from synthesized bulk black phosphorus crystal on 285-nm $SiO_2$ substrate.

**Transient reflection spectroscopy.**
To perform transient reflection spectroscopy measurement, a 250 kHz Ti-sapphire amplifier (RegA) system[47], an optical parametric amplifier (OPA), and a homemade difference frequency generator (DFG) were used. Laser pulses at 800-nm (1.55 eV) wavelength with 60-fs pulse width generated by RegA were injected into the OPA to generate 1350-nm (signal) and 1940-nm (idler) pulses. The 1940-nm pulse (~150 fs) was used as either pump in 1940-nm/4000-nm measurement or probe in 800-nm/1940-nm measurement. The dispersion-compensated residual 800-nm of the OPA was used as the pump. The OPA signal and idler can also be injected into the DFG to generate 4-μm pulse which is used as probe in 1940-nm pump experiment. The pump and probe are both linearly polarized and two half-wave plates are used to alter their polarization angle, respectively. A 40×tranmissive or reflective objective lens was used to focus the co-propagating pump and probe beams onto the sample that was placed in a liquid helium-cooled cryostat for temperature control. A liquid cryogen-free superconducting magnet system was used to generate a magnetic field up to 9 T perpendicular to the sample. The reflected probe beam was collected by the same reflective lens and detected by an InGaAs detector. The detected signal was read by lock-in amplifier referenced to 1.4-kHz mechanically chopped pump. The probe spot size is about 4 μm for 1940 nm probe and 8 μm for 4000 nm probe and the pump spot size is slightly larger.

**Acknowledgement**

This project has been supported by the National Basic Research Program of China (973 Grant No. 2014CB920900), the National Natural Science Foundation of China (NSFC Grant Nos. 11674013,11434010, 11704012), National Key Research and Development Program of China (Grant Nos: 2016YFA0300802, 2016YFE0110000, 2017YFA0303400), the Recruitment Program of Global Experts and the China Postdoctoral Science Foundation (Grant No:2017M610009).

**Author Contribution**

D.S. and K.C. conceived the idea and D.S. designed the experiments. C.Z. synthesized the BP bulk crystal under the supervision of S.J. X.L., L.W., J.L. and S.G. performed the transient reflection measurements under the supervision of D.S. X.Z. and Y.Z. performed the theoretical calculations under the supervision of K.C. X.L., W.L., X.Z., K.C. and D.S. analyzed the results, D.S. and K.C. wrote the manuscript, assisted by X.L., L.W., C.S.M., and X.Z. All the authors comments on the manuscript.

**Additional Information**

**Supplementary Information** accompanies this paper at http:/www.nature.com/naturecommunications
**Competing Interest:** The authors declare no competing financial interest

File Name: Supplementary Information

Description: Supplementary Notes, Supplementary Figures, Supplementary Table and Supplementary References.

## Supplementary Note 1. Dependence of $\Delta R/R|_{t=0}$ on pump polarization

As $\Delta R/R|_{t=0}(\alpha)$ is proportional to photoexcited carrier density as verified in pump power dependent measurement in reference[2], $\Delta R/R|_{t=0}(\alpha) \sim A(\alpha)$. The absorption coefficient $A$ of the black phosphorus (BP) crystal is given by[1, 2]:

$$A(\alpha) = \frac{4\varepsilon_0 c\sqrt{\varepsilon_1}(Re(R_{xx})cos^2(\alpha)+Re(\sigma_{yy})sin^2(\alpha))}{[\varepsilon_0 c(\sqrt{\varepsilon_1}+\sqrt{\varepsilon_2})+Re(\sigma_{xx})cos^2(\alpha)+Re(\sigma_{yy})sin^2(\alpha)]^2+[Im(\sigma_{xx})cos^2(\alpha)+Im(\sigma_{yy})sin^2(\alpha)]^2} \quad (1)$$

where, $\varepsilon$ is the permittivity of free space, $\varepsilon_1$ and $\varepsilon_2$ are the dielectric constants of air and $SiO_2$, respectively, in our sample geometry; $c$ is the speed of light and $\alpha$ is the angle of light polarization with respect to the armchair direction of BP crystal. In Supplementary Equation (1), $\sigma_0 = e^2/4\hbar$, $\varepsilon_0 c = \frac{\sigma_0}{0.023} = 43.5\sigma_0$, $Re(\sigma)$ and $Im(\sigma)$ are real and imaginary part of $\sigma$, respectively.

Under two band model, for unexcited state of multilayer BP sample[1], $Re(\sigma) \ll \varepsilon_0 c(\sqrt{\varepsilon_2}+\sqrt{\varepsilon_1})$ and $(Im(\sigma))^2 \ll [\varepsilon_0 c(\sqrt{\varepsilon_2}+\sqrt{\varepsilon_1})]^2$, so Supplementary Equation (1) can be simplified to the following:

$$A(\alpha) \approx \frac{4\sqrt{\varepsilon_1}(Re(\sigma_{xx})cos^2(\alpha)+Re(\sigma_{yy})sin^2(\alpha))}{\varepsilon_0 c(\sqrt{\varepsilon_2}+\sqrt{\varepsilon_1})^2} \quad (2)$$

which is proportional to the real part of complex optical conductivity.

## Supplementary Note 2. Faraday rotation of the transmissive objective lens under high magnetic field.

Unlike reflective objective lens, the light passing through transmissive objective lens experience a Faraday rotation due to the effect of transmissible materials present in the lens. Hence, while measuring the effect of polarization using transmissive objective lens, one should take into account the rotation induced by Faraday effects.

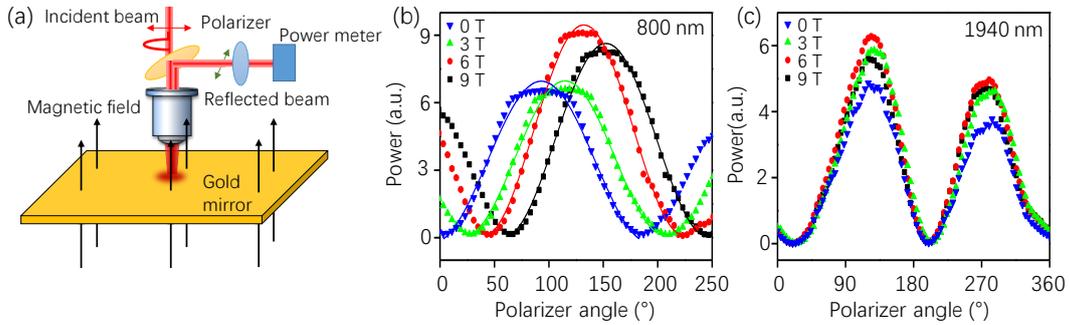

**Supplementary Figure 1. a,** Schematic diagram of the optical setup used to measure the magnetic field induced change in polarization, **b, c.** Polarization analysis of Faraday rotation of 800-nm and 1940-nm beams under different magnetic field. x-axis is the polarization angle of analytic polarizer and y-axis is the transmitted power (measured by power meter) through analytic polarizer.

Faraday rotation of light is determined by measuring the pump (800-nm) and probe (1940-nm) polarization rotation using the optical setup shown in Supplementary Fig. 1a. In these measurements, the incident beam with fixed polarization was focused on to a gold mirror in normal incidence geometry by transmissive objective lens and the reflected beam was collected by the same objective lens. The reflected beam was then sampled by a partial reflector and its polarization was identified by an analytic polarizer together with a power meter. Supplementary Fig. 1b & c show the reflected power as a function of polarization angle of analytic polarizer. As the magnetic field increases from 0 to 9 T, polarization of pump beam rotates up to 50°, but the polarization of probe beam remains constant indicating that only 800-nm beam experiences significant Faraday rotation while passing through the objective lens. As the reflected 800-nm beam passes through the objective lens (Supplementary Fig. 1b) twice during the measurement, its single-pass Faraday rotation angle can be determined from the change in the polarization angle of pump beam under the applied magnetic field. The variation in the pump beam polarization angle with increasing magnetic field is shown as red circles in Supplementary Fig. 2b and the Faraday rotation angle can be fitted with $\varphi_B$=3.22·B(degree).

The Faraday rotation of the pump polarization is further confirmed from the $\Delta R/R|_{t=0}$ signal recorded as a function of incident pump polarization with fixed probe polarization at 0°, which is shown in Supplementary Fig. 2a. It is clearly evident from this figure that the ellipse rotates with respect to applied magnetic field. The pump polarization dependent $\Delta R/R|_{t=0}$ signal is fitted with the following function:

$$\Delta R/R|_{t=0}(\alpha_1) = A_1 cos^2(\alpha_1+\varphi+\varphi_B) + A_2 sin^2(\alpha_1+\varphi+\varphi_B) \quad (3)$$

where $\alpha_1$ is the polarization angle of incident light, $\varphi$ (= -9.3°) is the angle between crystal x-axis and initially arbitrarily defined 0° of pump polarization, and $\varphi_B(B)$ is the magnetic field dependent rotation angle of the ellipse. The fitting parameters $A_1$ and $A_2$ correspond to the long and short axes of ellipse, respectively and their ratio is given by $A_1/A_2 = Re(\sigma_{xx})/Re(\sigma_{yy})$, where $Re(\sigma_{xx})$ and $Re(\sigma_{yy})$ is the real part of optical conductivity of BP at 800 nm. The obtained rotation angle $\varphi_B$ has linear dependence on the applied magnetic field B (represented as black circles in Supplementary Fig. 2b) with slope = 2.78. As shown in Supplementary Fig. 2b, considering the error bar, $\varphi_B$ matches with $\varphi_{pump}$, determined from Supplementary Fig. 1b, confirming that the rotation of conductivity ellipse is solely due to the Faraday rotation of pump polarization when passing through the transmissive objective lens. Furthermore, it should be taken into account that this rotation is not observed when a reflective objective lens is used.

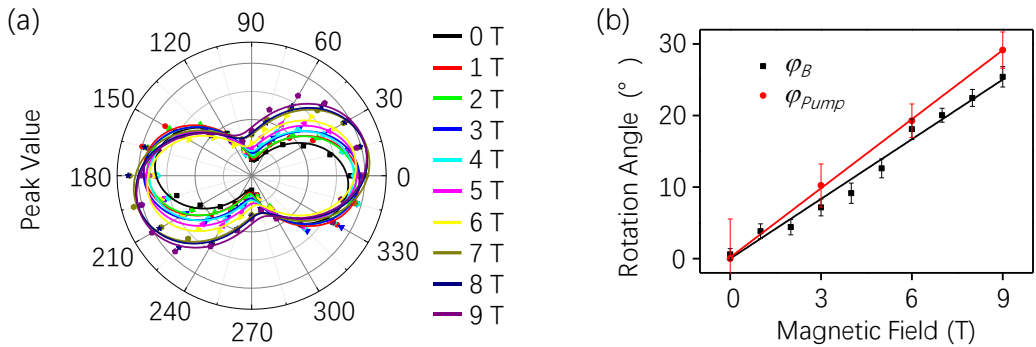

**Supplementary Figure 2. Pump polarization dependence of $\Delta R/R|_{t=0}$ under different magnetic field a,** $\Delta R/R|_{t=0}$ as a function of pump polarization with fixed probe polarization at 0° under different magnetic fields. **b,** Comparison of the results of $\varphi_B$ (black squares) and $\varphi_{pump}$ (red circles) obtained from Supplementary Fig. 2a and 1b, respectively. Solid lines are the linear fits of $\varphi_B$ and $\varphi_{pump}$ as a function of applied magnetic field.

## Supplementary Note 3. 800-nm pump polarization dependent results

Supplementary Fig. 3 shows the 800-nm pump polarization dependent measurements under different magnetic field up to 9 T. The full time scans of the carrier dynamics at different pump polarization and fixed probe polarization are shown in Supplementary Fig. 3a. The dependence of transient reflection $\Delta R/R$ amplitude on pump polarization is observed while their relaxation dynamics remains the same. Supplementary Fig. 3b shows the evolution of experimentally measured conductivity ellipses as function of applied magnetic field. An increase in the magnitude of both $Re(\sigma_{xx})$ and $Re(\sigma_{yy})$ is observed with increasing magnetic field, but the relative increment of $Re(\sigma(B))/Re(\sigma(0))$ is much larger in y-direction compared to that in x-direction (Supplementary Fig. 3c), indicating the anisotropic response(AR) of BP degrades as magnetic field increases. As shown in Supplementary Fig. 3d, the ratio of $Re(\sigma_{xx})/Re(\sigma_{yy})|_{800\ nm}$, which describes the AR of BP at 800 nm, decreases from 6.8 to 2.7 when the magnetic field is increased from 0 to 9 T.

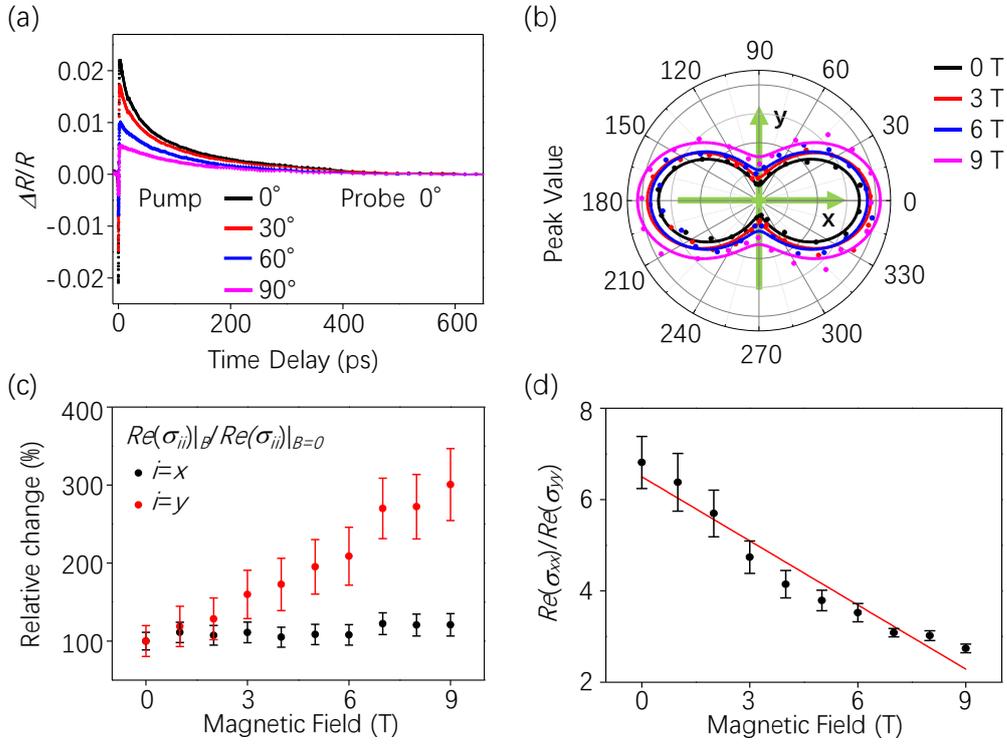

**Supplementary Figure 3. Pump polarization dependence of transient reflection spectra under magnetic field a,** Transient reflection spectra with 800-nm pump polarization fixed at 0°, 30°, 60° and 90° with respect to x-axis and 1940-nm probe polarization fixed along x-axis. The solid lines are biexponential fittings. **b,** $\Delta R/R|_{t=0}$ as a function of pump polarization with different magnetic field at 0, 3, 6 and 9 T respectively. The polarization dependence of $\Delta R/R|_{t=0}$ is fitted by the function $A_1 cos^2(\alpha_1+\varphi_B)+A_2 sin^2(\alpha_1+\varphi_B)$. The data has been rotated $\varphi_B$ degree to remove the Faraday rotation effect due to the transmissive objective lens, so that the long axis of the fitting curve can be parallel with the x-axis of BP. **c,** The relative change of real part of conductivity $Re(\sigma_{ii})|_B/Re(\sigma_{ii})|_{B=0}$, $i=x$ or $y$. The black squares and red circles with error bars are conductivity change along x- and y-axis, respectively. **d,** The anisotropy of conductivity ellipse: $Re(\sigma_{xx})/Re(\sigma_{yy})$ as function of magnetic field. The solid line is a linear fit. All the experiments are performed at 78 K.

**Supplementary Note 4. Pump induced conductivity changes**

When the probe polarization is rotated, the shape as well as the magnitude of the transient reflection spectra changes (Supplementary Fig. 4a). In all probe polarization dependent measurements, a reflective objective lens is used to avoid the Faraday rotation. Under two band model and using similar approximations used in the reference[1]: $Im^2(\sigma)<<\varepsilon_0^2 c^2$, $Re(\sigma)<<\varepsilon_0 c$, and $\Delta\sigma<<\sigma$, which is valid at quasi-equilibrium states of photoexcited carriers after the initial thermalization and cooling from highly unequilibrium state, the $\Delta R/R$ can be expressed as a function of conductivity changes due to the pump excitation, $\Delta Re(\sigma_{xx})$ and $\Delta Re(\sigma_{yy})$[2]:

$$\frac{\Delta R}{R}|_t(\alpha_2) \approx \frac{0.03}{\sigma_0}(\Delta Re(\sigma_{xx})\cos^2\alpha_2 + \Delta Re(\sigma_{yy})\sin^2\alpha_2) \qquad (4)$$

Where $\alpha_2$ is the polarization angle of the probe with respect to x-axis of BP. The experimentally measured probe polarization dependence of $\Delta R/R$ signal at certain delay time can be fitted with:

$$\frac{\Delta R}{R}|_t = A_1\cos^2\alpha_2 + A_2\sin^2\alpha_2 \qquad (5)$$

which is shown in Supplementary Fig. 4b. From Supplementary Equation (4) and (5), the conductivity change $\Delta Re(\sigma_{xx})|_t$ and $\Delta Re(\sigma_{yy})|_t$ at certain delay time ($t$) can be directly obtained from the fitting parameters $A_{1/2}$ from the following relation:

$$\Delta Re(\sigma_{xx})|_t = \frac{\sigma_0}{0.03}A_1, \quad \Delta Re(\sigma_{yy})|_t = \frac{\sigma_0}{0.03}A_2 \qquad (6)$$

According to Supplementary Equation (6), $\Delta Re(\sigma_{xx,yy})|_t$ can be retrieved from the probe polarization dependent experiments. The same method also applies to $\Delta Re(\sigma_{xx/yy})|_t(B)$.

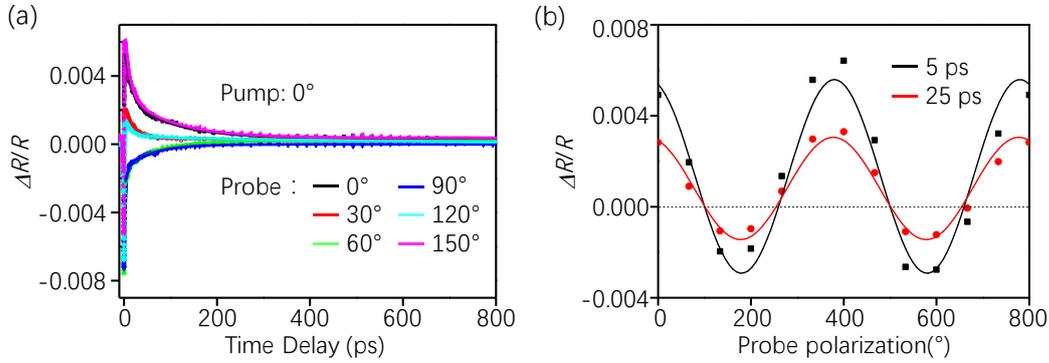

**Supplementary Figure 4. Probe polarization dependence of transient reflection spectra. a,** Probe polarization dependence of transient reflection spectra with pump polarization fixed along crystal x-axis. The pump wavelength is 800 nm and the probe wavelength is 1940 nm. The decay dynamics are fitted by biexponential function; **b,** $\Delta R/R$ signal as a function of probe polarization angle $\alpha_2$ at 5 ps and 25 ps delay times.

**Supplementary Note 5. Polarization dependent reflection under magnetic field**

The pump polarization dependent transient reflection measurement with 1940-nm pump can provide $Re(\sigma_{xx})/Re(\sigma_{yy})$ at different magnetic field. To get the absolute value of $Re(\sigma_{xx/yy})$, we tried to retrieve $Re(\sigma_{xx})-Re(\sigma_{yy})$ from polarization dependent transient reflection measurement of BP under magnetic field.

The reflectivity of BP crystal is given by[2]:

$$R \approx \left|\frac{\varepsilon_0 c(\sqrt{\varepsilon_2}-\sqrt{\varepsilon_1})+Re(\sigma_{xx})cos^2(\alpha)+Re(\sigma_{yy})sin^2(\alpha)}{\varepsilon_0 c(\sqrt{\varepsilon_2}+\sqrt{\varepsilon_1})+Re(\sigma_{xx})cos^2(\alpha)+Re(\sigma_{yy})sin^2(\alpha)}\right|^2 = \left|\frac{\varepsilon_0 c\left(\sqrt{\varepsilon_2}+\frac{Re(\sigma_{yy})}{\varepsilon_0 c}-1\right)+(Re(\sigma_{xx})-Re(\sigma_{yy}))cos^2(\alpha)}{\varepsilon_0 c\left(\sqrt{\varepsilon_2}+\frac{Re(\sigma_{yy})}{\varepsilon_0 c}+1\right)+(Re(\sigma_{xx})-Re(\sigma_{yy}))cos^2(\alpha)}\right|^2 \quad (7)$$

where $\varepsilon_0$ is the free-space permittivity, $\varepsilon_1$ and $\varepsilon_2$ are the dielectric constants of air and substrate, respectively, in our sample geometry; $c$ is the speed of light, and $\alpha$ is the light polarization angle with respect to crystal x-axis.

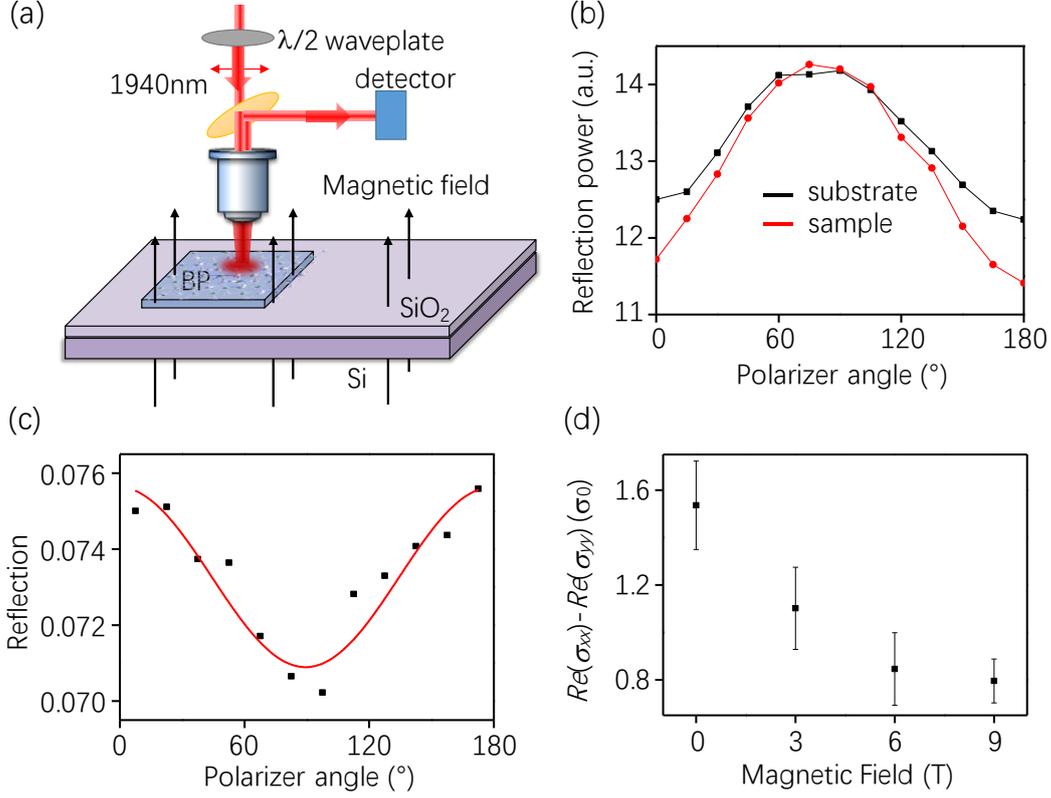

**Supplementary Figure 5. Probe polarization dependent reflection. a,** Schematic diagram of the setup used to measure the polarization dependent reflection. **b,** Relative reflection of BP sample and bare substrate. **c,** Reflection of BP and its fitting curve. The polarizer angle has been changed according to the fitting parameter $\alpha$, so that the 0° and 90° correspond to the x- and y-axis of the sample. **d,** $Re(\sigma_{xx})-Re(\sigma_{yy})|_{1940\ nm}$ under different magnetic fields.

Supplementary Fig. 5a shows the schematic diagram of the setup used to measure polarization resolved reflection. The polarization direction of 1940-nm beam was controlled by a half-wave plate before it passes through the objective lens. Here, a reflective objective lens was used to focus the 1940-nm beam onto the sample to avoid Faraday rotation. Due to the AR of polarization sensitive optical elements in the setup, such as beam splitter, photo detector et al., the direct measurement of the BP sample reflection is not accurate. In order to rule out these extrinsic effects, the power of reflective beam of a bare substrate $I_{sub}$ was measured beforehand to calibrate the polarization dependent response of the system. Then the focal spot was moved onto the sample and the reflected power $I_{BP}$ from BP sample was measured with the same experimental conditions, as shown in Supplementary Fig. 5b.

The reflection of BP sample then can be expressed as:

$$R_{BP}(\alpha) = \frac{I_{BP}(\alpha)}{I_{sub}(\alpha)} R_{sub} \qquad (8)$$

where, $I_{BP}$ and $I_{sub}$ are the intensities of reflected light with polarization angle $\alpha$ from BP sample and bare substrate, respectively. As shown in Supplementary Fig. 5c, $R_{BP}(\alpha)$ can be fitted by:

$$R_{BP}(\alpha) = \left|\frac{\varepsilon_0 c(a-1)+b\cos^2(\alpha)}{\varepsilon_0 c(a+1)+b\cos^2(\alpha)}\right|^2 \qquad (9)$$

where, $a=\sqrt{\varepsilon_2} + Re(\sigma_{yy})/\varepsilon_0 c$, with $\sqrt{\varepsilon_2} = n_{sub}$, $\varepsilon_2$ and $n_{sub}$ are defined as dielectric constant and refractive index of the substrate, respectively. $b=Re(\sigma_{xx})-Re(\sigma_{yy})$ and $\alpha$ is the polarization angle of 1940 nm.

Here $Re(\sigma_{yy})/\varepsilon_0 c \sim 0.005 \ll \sqrt{\varepsilon_2}$, so it is not reliable to get $Re(\sigma_{yy})$ directly from fitting parameter $a$, because a small deviation in $A$ can introduce a large error on $Re(\sigma_{yy})$. However, $b= Re(\sigma_{xx})-Re(\sigma_{yy})$ can be reliably obtained from the fitting of $R(\alpha)$ as shown in Supplementary Table 1.

As the substrate of BP is 285 nm SiO$_2$ on Si, it is helpful to deduct the effective refractive index $\varepsilon_2$ of the substrate. Due to its isotropic lattice structure, the substrate has isotropic reflective response to normal incident light, and hence the reflection of substrate does not have any polarization dependence. Characteristic transfer matrix of SiO$_2$ that is used in the reflection calculation is:

$$M = \begin{bmatrix} \cos\delta_1 & -\frac{i\sin\delta_1}{\eta_1} \\ -i\eta_1\sin\delta_1 & \cos\delta_1 \end{bmatrix} \qquad (10)$$

where, $\delta_1 = \frac{2t}{\lambda}n_1 d_1$ is the phase delay induced by the SiO$_2$ thin film, $\eta_1 = \sqrt{\frac{\varepsilon_0}{\mu_0}}n_1$ is the effective optical admittance for normal incidence. Using thickness of SiO$_2$ ($d_1 = 285$ nm), refractive indices of air ($n_0 = 1$), SiO$_2$ ($n_1 = 1.4376$)[3] and Si ($n_2 = 3.4263$)[4] for 1940-nm wavelength at 77 K and effective optical admittance of Si: $\eta_2 = \sqrt{\frac{\varepsilon_0}{\mu_0}}n_2$, the reflection of substrate can be obtained through the following equation:

$$R_{sub} = \frac{(\eta_0-\eta_2)^2\cos^2\delta_1+(\frac{\eta_0\eta_2}{n_1}-\eta_1)^2\sin^2\delta_1}{(\eta_0+\eta_2)^2\cos^2\delta_1+(\frac{\eta_0\eta_2}{n_1}+\eta_1)^2\sin^2\delta_1} = 0.0794 \qquad (11)$$

If we consider the whole substrate Si/SiO$_2$ together, the 1940-nm light would see a substrate with equivalent reflective index of $n_{sub}=1.7847$. As shown in Supplementary Table 1, the fitting results of $A = n_{sub} + Re(\sigma_{yy})/\varepsilon_0 c$ are within reasonable agreement with the equivalent reflective index of the substrate.

**Supplementary Table 1.** Polarization dependent reflection fitting parameters of $a$, $b$

| Magnetic Field | a | | b | |
|---|---|---|---|---|
| | Value | Error | Value | Error |
| 0T | 1.72575 | 0.00236 | 1.43162 | 0.17383 |
| 3T | 1.73786 | 0.00218 | 1.02633 | 0.16098 |
| 6T | 1.72138 | 0.00193 | 0.78859 | 0.14233 |
| 9T | 1.74744 | 0.00129 | 0.73545 | 0.08633 |

Using the above value of $b=Re(\sigma_{xx})-Re(\sigma_{yy})$ at different magnetic field determined from the polarization dependent reflection measurement at 1940 nm, and using the conductivity ratio $Re(\sigma_{xx})/Re(\sigma_{yy})|_{1940\,nm}$ measured from the 1940-nm pump polarization dependent experiments as shown in Fig. 2, the absolute values of $Re(\sigma_{xx})|_{1940\,nm}$ and $Re(\sigma_{yy})|_{1940\,nm}$ can be obtained.

**Supplementary Note 6. Hot carrier induced anisotropy change under magnetic field**

With the conductivity value $Re(\sigma_{xx})|_{1940\,nm}$, $Re(\sigma_{yy})|_{1940\,nm}$ extracted from pump polarization dependent transient reflection measurement discussed above and conductivity change $\Delta Re(\sigma_{xx})|_{1940\,nm}$, $\Delta Re(\sigma_{yy})|_{1940\,nm}$ extracted from probe polarization dependent measurement, the hot carrier induced conductivity anisotropy change at certain magnetic field: $\Delta(B)$ can be calculated through the following definition:

$$\Delta(B) = \frac{Re(\sigma_{xx})+\Delta Re(\sigma_{xx})}{Re(\sigma_{yy})+\Delta Re(\sigma_{yy})}|_B - \frac{Re(\sigma_{xx})}{Re(\sigma_{yy})}|_B = \frac{\Delta Re(\sigma_{xx})Re(\sigma_{yy})-Re(\sigma_{xx})\Delta Re(\sigma_{yy})}{Re(\sigma_{yy})*(Re(\sigma_{yy})+\Delta Re(\sigma_{yy}))}|_B \quad (12)$$

The relative change of $\Delta$ can be written as:

$$\Delta\%(B) = \left(\frac{Re(\sigma_{xx})+\Delta Re(\sigma_{xx})}{Re(\sigma_{yy})+\Delta Re(\sigma_{yy})}|_B - \frac{Re(\sigma_{xx})}{Re(\sigma_{yy})}|_B\right)/\frac{Re(\sigma_{xx})}{Re(\sigma_{yy})}|_B = \frac{\Delta Re(\sigma_{xx})Re(\sigma_{yy})-\Delta Re(\sigma_{yy})Re(\sigma_{xx})}{Re(\sigma_{xx})Re(\sigma_{yy})+\Delta Re(\sigma_{yy})Re(\sigma_{xx})}|_B \quad (13)$$

**Supplementary Note 7. Measurement with fine tuning of magnetic field**

Supplementary Fig. 6 shows the evolution of AR of BP when magnetic field is tuned around 7 T with fine steps down to 0.01 T. The evolution of experimentally measured conductivity ellipses as function of magnetic fields are shown in Supplementary Fig. 6a, the ratio of $Re(\sigma_{xx})/Re(\sigma_{yy})$ has small variation when the magnetic field increases from 7.0 to 7.3 T (Supplementary Fig. 6b). It is not clear from these measurements whether the small fluctuations of $Re(\sigma_{xx})/Re(\sigma_{yy})$ are oscillations due to LLs, predicted by theory shown in Fig. 4b, as these fluctuations are within the experimental uncertainty and it is impossible to match the oscillation period with the theoretical predictions.

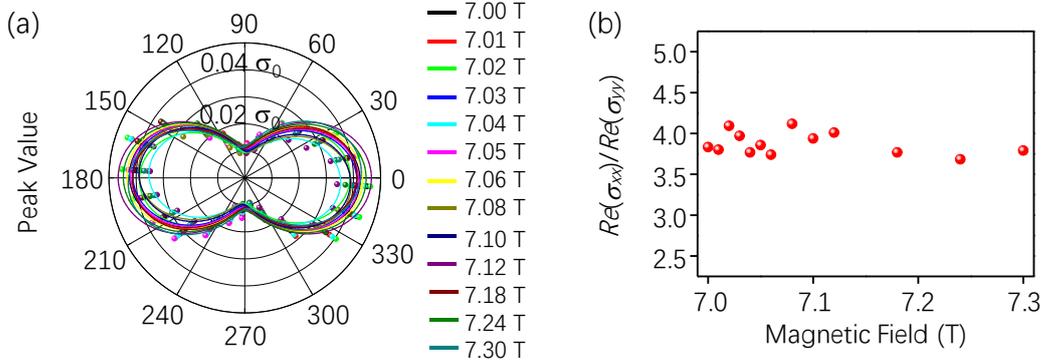

**Supplementary Figure 6. Fine tuning of magnetic field a,** Conductivity ellipse ($\Delta R/R|_{t=0}$ as function of pump polarization) with magnetic field increasing from 7.0 to 7.3 T. The polarization dependence of $\Delta R/R|_{t=0}$ is fitted by the function $A_1 cos^2(\alpha+\varphi_B)+A_2 sin^2(\alpha+\varphi_B)$. The data is rotated $\varphi_B$ degree to account the Faraday rotation so that the long axis of the fitting curve can be parallel with the x-axis of BP. **b,** The anisotropy of conductivity ellipse: $Re(\sigma_{xx})/Re(\sigma_{yy})$ as a function of magnetic field.

**Supplementary Note 8. Calculation of Magneto-optical conductivity and Landau level wavefunction of BP.**

The low-energy physics of bulk BP around the Z point can be well described by an anisotropic two-band $k \cdot p$ Hamiltonian, which is given by

$$H = \begin{pmatrix} h_c & h_{cv} \\ h_{vc} & h_v \end{pmatrix} = \begin{pmatrix} E_c + ck_x^2 + ck_y^2 + \delta_c k_z^2 & \gamma k_x \\ \gamma k_x & E_v - \lambda k_x^2 - \eta k_y^2 - \delta_v k_z^2 \end{pmatrix} \tag{14}$$

where $E_c = E_g/2, E_v = -E_g/2$ with $E_g$ the band gap. Recent results[5] show that the band gap is 0.35 eV. Other parameters are[6] : $\alpha = 0.253$ eV·nm$^2, \beta = 0.036$ eV·nm$^2, \lambda = 0.313$ eV·nm$^2, \eta = 0.054$ eV·nm$^2, \delta_c = 0.131$ eV·nm$^2, \delta_v = 0.063$ eV·nm$^2$, $\gamma = 0.23$ eV·nm.

When coupled with a perpendicular magnetic field B = (0,0,B), taking the Landau gauge A = (−By, 0,0), we can define the creation and annihilation operators as

$$a = \sqrt{\frac{m_{cy}\omega_c}{2\hbar}}\left(y - y_0 + \frac{ip_y}{m_{cy}\omega_c}\right), a^\dagger = \sqrt{\frac{m_{cy}\omega_c}{2\hbar}}\left(y - y_0 - \frac{ip_y}{m_{cy}\omega_c}\right) \tag{15}$$

where $\omega_c = eB/\sqrt{m_{cx}m_{cy}}$ is the cyclotron frequency with $m_{cx} = \hbar^2/2\alpha$, $m_{cy} = \hbar^2/2\beta$, $y_0 = k_x l_B^2$ is the cyclotron center, $l_B = \sqrt{\hbar/eB}$ is the magnetic length. Therefore, the Hamiltonian can be rewritten as

$$H = \begin{pmatrix} E_c + \hbar\omega_c\left(a^\dagger a + \frac{1}{2}\right) + \delta_c k_z^2 & \hbar\omega_\gamma(a^\dagger + a) \\ \hbar\omega_\gamma(a^\dagger + a) & E_v - \hbar\omega_v\left(a^\dagger a + \frac{1}{2}\right) - \hbar\omega'\left(a^2 + a^{\dagger 2}\right) - \delta_v k_z^2 \end{pmatrix} \tag{16}$$

where, $\omega_\gamma = \gamma/\sqrt{2}\hbar l_B \alpha_{yx}$, $\omega_v = (r_x + r_y)\omega_c, \omega' = (r_x - r_y)\omega_c$ with $\alpha_{yx} = (m_{cy}/m_{cx})^{\frac{1}{4}}$, $r_x = m_{cx}/m_{vx}, r_y = m_{cy}/m_{vy}, m_{vx} = \hbar^2/2\lambda$, $m_{vy} = \hbar^2/2\eta$. The wavefunction of the system can be expressed as

$$\psi = \sum_m \begin{pmatrix} c_m \\ d_m \end{pmatrix} |m, k_x\rangle \tag{17}$$

where, $|m, k_x\rangle$ is the eigenvectors of the number operator $n = a^\dagger a$. Then, the Schrödinger equation $H\psi = E_{n,k_z}\psi$ can be solved numerically and obtain the eigenvalues as well as the eigenvectors. Considering the Landau level broadening arising from the random potential caused by defects and impurities in actual samples, a Gaussian profile is used for the density of states, which is given by

$$D(\varepsilon) = g_s \sum_{n,k_x,k_z} \frac{1}{\sqrt{2\pi}\Gamma} exp\left[-\frac{(\varepsilon - E_{n,k_z})^2}{2\Gamma^2}\right] \tag{18}$$

where $E_{n,k_z}$ is the eigenvalues of the Hamiltonian, $\Gamma$ is the Landau level broadening, $g_s = 2.0$ for the spin degeneracy. Hence, the Fermi level is defined by $\int d\epsilon [1 - f(\epsilon)]D(\epsilon) = n_h$ with $f(\epsilon)$ is the Fermi-Dirac distribution and $n_h$ is the hole concentration. Following our previous work[7], the longitudinal optical-conductivity calculated via Kubo-Greenwood formula is

$$\sigma_{xx/yy} = \frac{e^2 \hbar}{iS_0} \sum_{\xi \neq \xi'} \int \frac{dk_z}{2\pi} \frac{f(E_\xi) - f(E_{\xi'})}{\left(E_\xi - E_{\xi'}\right)} \frac{|\langle \xi | v_{x/y} | \xi' \rangle|^2}{\left(E_\xi - E_{\xi'} + \hbar\omega + i\Gamma\right)} \tag{19}$$

with the velocity matrices

$$v_x = \begin{pmatrix} -v_1(a^\dagger + a) & v_f \\ v_f & r'_x v_1(a^\dagger + a) \end{pmatrix}, v_y = \begin{pmatrix} -1 & 0 \\ 0 & r'_y \end{pmatrix} iv_2(a^\dagger - a),$$

where, $\hbar\omega$ is the photon energy, $|\xi\rangle = |s, n, k_x, k_z\rangle$ are the eigenstates of the system, the velocities are $v_f = 3.5 \times 10^5$ m/s, $v_1 = \sqrt{\hbar\omega_c/2m_{cx}}, v_2 = \sqrt{\hbar\omega_c/2m_{cy}}$, $r'_x = 2r_x, r'_y = 2r_y$.

From the numerically obtained eigenvalue and eigenvectors, the value of magneto-conductivity is determined from Supplementary Equation (19).

As shown in Supplementary Fig.7, the ratio of $Re(\sigma_{xx})/Re(\sigma_{yy})$ is found to have temperature and doping dependence. However, the monotonic decrease of $Re(\sigma_{xx})/Re(\sigma_{yy})$ with magnetic field (B) is unaffected by the doping and temperature. Both temperature and doping only affect the absolute amplitude of $Re(\sigma_{xx})/Re(\sigma_{yy})$.

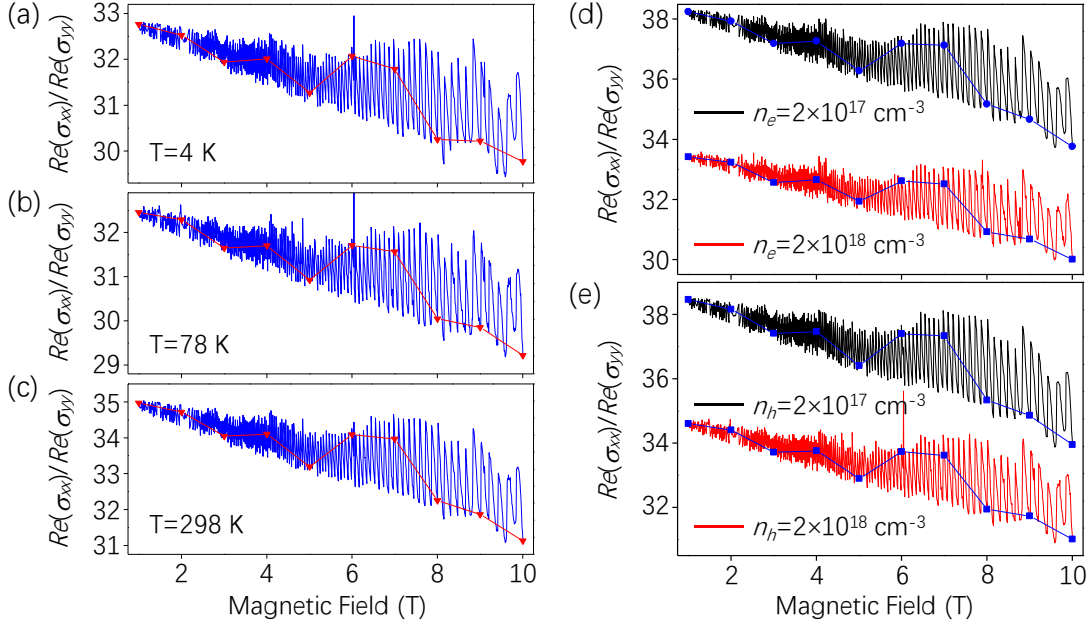

**Supplementary Figure 7. The magneto-conductivity ratio as a function of magnetic field B. a-c**, The conductivity ratio $Re(\sigma_{xx})/Re(\sigma_{yy})$ at temperature of 4, 78 and 298 K, respectively for hole doping concentration $n_h$=4.2×10$^{18}$ cm$^{-3}$ and **d-e**, for temperature T=78 K under different doping concentrations. The photon energy $\hbar\omega$ adopted is 0.64 eV and the Landau level broadening $\Gamma$ is 4 meV.

**Landau level Wavefunctions in symmetric gauge of BP**

To see the anisotropy of wavefunctions under magnetic field, we adopt the symmetric gauge $B = (-By, Bx, 0)/2$. Similarly, we choose the eigenstates of $h_c$ as a basis to obtain the Landau levels of the whole system. To diagonalize $h_c$, firstly, we should perform a coordinate transformation

$$u = \frac{1}{\sqrt{2\chi}}(x + \chi y), v = \frac{1}{\sqrt{2\chi}}(x - \chi y) \tag{20}$$

where $\chi = \sqrt{m_{cy}/m_{cx}}$. Then, we have

$$p_u = \frac{1}{\sqrt{2\chi}}(p_y + \chi p_x), p_v = \frac{1}{\sqrt{2\chi}}(p_y - \chi p_x) \tag{21}$$

One can easily verify that $[u, p_u] = i\hbar$, $[v, p_v] = i\hbar$. When a perpendicular magnetic field applied, the $h_c$ in symmetric gauge can be written as

$$h_c = \frac{1}{2\sqrt{m_{cx}m_{cy}}}[(p_u - eBv/2)^2 + (p_v + eBu/2)^2].$$

Defining the following operators

$$c = \frac{1}{\sqrt{2}l_B}\left[\left(-\frac{v}{2} + l_B^2 k_u\right) - i\left(\frac{u}{2} + l_B^2 k_v\right)\right], c^\dagger = \frac{1}{\sqrt{2}l_B}\left[\left(-\frac{v}{2} + l_B^2 k_u\right) + i\left(\frac{u}{2} + l_B^2 k_v\right)\right],$$

$$b = \frac{1}{\sqrt{2}l_B}\left[\left(\frac{u}{2} - l_B^2 k_u\right) + i\left(\frac{u}{2} + l_B^2 k_v\right)\right], b^\dagger = \frac{1}{\sqrt{2}l_B}\left[\left(\frac{u}{2} - l_B^2 k_u\right) - i\left(\frac{u}{2} + l_B^2 k_v\right)\right] \tag{22}$$

we find that $h_c = \left(m + \frac{1}{2}\right)\hbar\omega_c$. Similarly, the Hamiltonian of BP can be rewritten as

$$H = \begin{pmatrix} E_c + \hbar\omega_c\left(b^\dagger b + \frac{1}{2}\right) + \delta_c k_z^2 & \frac{-\gamma}{2\sqrt{\chi}l_B}[(1+i)b + (1-i)b^\dagger] \\ \frac{-\gamma}{2\sqrt{\chi}l_B}[(1-i)b^\dagger + (1+i)b] & E_v - \left(b^\dagger b + \frac{1}{2}\right)\hbar\omega_{v1} - i\left(b^2 - b^{\dagger 2}\right)\hbar\omega_{v2} - \delta_v k_z^2 \end{pmatrix} \tag{23}$$

where $\omega_{v1} = (\omega_{vx} + \omega_{vy})/2, \omega_{v2} = (\omega_{vx} - \omega_{vy})/4$, with $\omega_{vx} = 2\lambda/\hbar l_B^2 \chi$, $\omega_{vy} = 2\chi\eta/\hbar l_B^2$. The corresponding eigenvector is

$$|n,m\rangle = \psi_{n,m}(x,y) = A_{n,m} e^{i(n-m)\theta} e^{-|r|^2/4} r^{|m-n|} L_{(n+m-|m-n|)/2}^{|m-n|}(|r|^2/2) \tag{24}$$

where $A_{n,m}$ is a normalization constant, $r = u + iv$, $L_n^\alpha(x)$ is the associated Laguerre polynomials. On the other hand, the operators satisfy $c|n,m\rangle = \sqrt{n}|n-1,m\rangle, c^\dagger|n,m\rangle = \sqrt{n+1}|n+1,m\rangle, b|n,m\rangle = \sqrt{m}|n,m-1\rangle, b^\dagger|n,m\rangle = \sqrt{m+1}|n,m+1\rangle$. Since the Hamiltonian is independent of operator $c$, the wavefunction of $n = 0$ case can be used as a basis to diagonalize the Hamiltonian. The wavefunction of the system is

$$\Phi(x,y,z) = \frac{\exp(ik_z z)}{\sqrt{L_z}} \sum_m \begin{pmatrix} c_m \\ d_m \end{pmatrix} \psi_{0,m}(x,y), \tag{25}$$

where $\psi_{0,m}(x,y) = \frac{1}{\sqrt{2,2^m m!}l_B}\left(\frac{r^*}{l_B}\right)^m \exp\left(-\frac{|r|^2}{4l_B}\right)$. Therefore, the spatial distribution of the probability density of the $n$-th Landau level is

$$|\Phi_n(x,y,z)|^2 = \sum_{j,k}\left(c_j^{n*}c_k^n + d_j^{n*}d_k^n\right)\psi_{0,j}(x,y)\psi_{0,k}(x,y) \tag{26}$$

Supplementary Fig. 8 shows the spatial distribution probability of the Landau level around the transition energy (0.64 eV) in the valence band for $k_z$=0 under magnetic field of 6 T and 9 T respectively. From Supplementary Fig. 8a&b, we find the cyclotron orbits under magnetic field are all ellipses. As shown in the normalized plot of Supplementary Fig. 8c, the elliptical wavefunction under magnetic field 6T is more anisotropic than that under 9T.

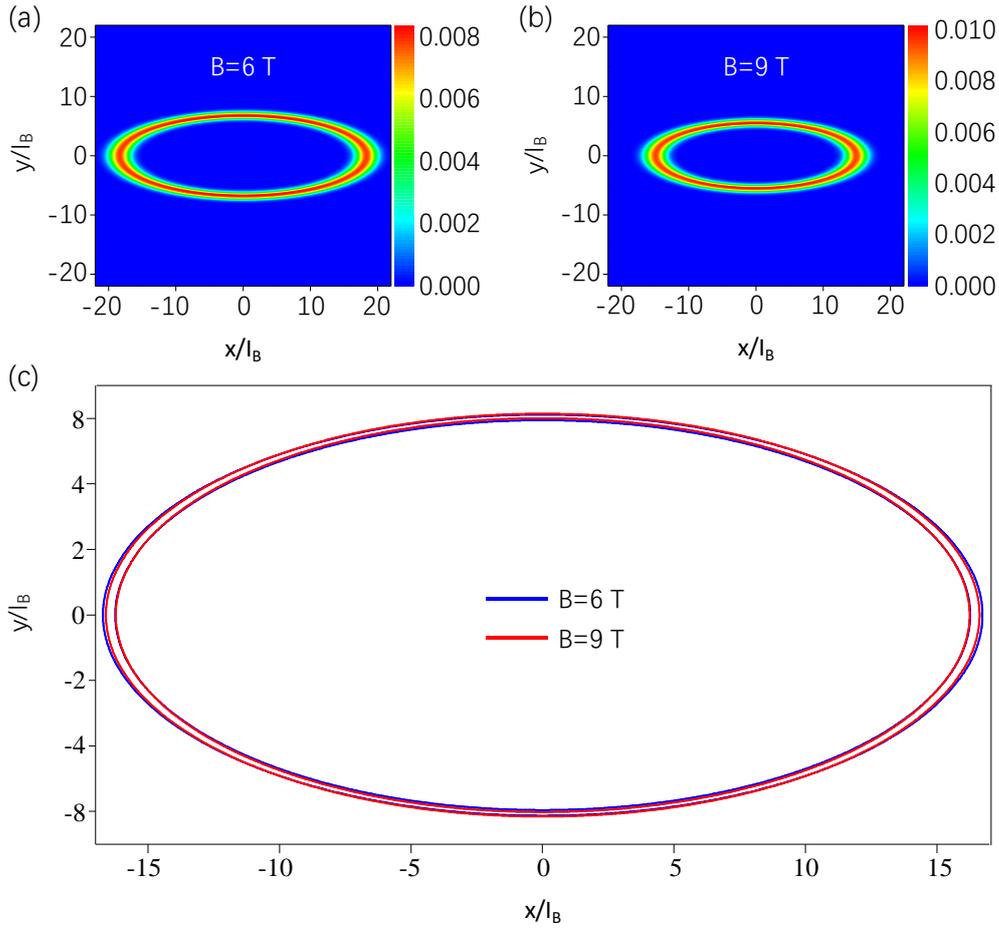

**Supplementary Figure 8. The spatial distribution density of the Landau level around transition energy in the valence band.** The spatial distribution of wavefunction of the Landau level around transition energy (0.64 eV) in the valence band for $k_z = 0$ under magnetic field **a,** $B = 6$ T and **b,** $B = 9$ T. **c,** Contour plot of the wavefunctions corresponding to a and b. As the contour lines of wavefunctions are concentric ellipses for a given magnetic field, to compare the change of anisotropy under different spring fields, two ellipses are chosen among all the contour lines with either identical long axis (see the two inner ellipses) or short axis (see the two outer ellipses).

**Supplementary References**